\documentclass[fleqn,usenatbib,useAMS]{mnras}

\usepackage{graphicx}	
\usepackage{amsmath}	
\usepackage{amssymb}	
\usepackage{multicol}   
\usepackage{bm}		    
\usepackage{pdflscape}	
\usepackage[dvipsnames]{xcolor}

\usepackage{tabularx}
\usepackage{threeparttable}

\usepackage[T1]{fontenc}
\usepackage{ae,aecompl}

\usepackage{newtxtext,newtxmath}

\title[The Common Fundamental Plane of Pulsars and Magnetars]{The Common Fundamental Plane of X-ray Emissions from Pulsars and Magnetars in Quiescence}

\author[C.-Y.\ Chu et al.]{Che-Yen Chu,$^{1}$\thanks{Contact e-mail: \href{mailto:cychu@gapp.nthu.edu.tw}{cychu@gapp.nthu.edu.tw}}
Hsiang-Kuang Chang$^{1,2}$
\\
$^{1}$Institute of Astronomy, National Tsing Hua University, Hisnchu 300044, Taiwan \\
$^{2}$Department of Physics, National Tsing Hua University, Hisnchu 300044, Taiwan }

\date{Accepted XXX. Received YYY.}

\pubyear{2023}

\begin{document}
\label{firstpage}
\pagerange{\pageref{firstpage}--\pageref{lastpage}}
\maketitle

\begin{abstract}
Magnetars are a unique class of neutron stars characterized by their incredibly strong magnetic fields. Unlike normal pulsars whose X-ray emission was driven by rotational energy loss, magnetars exhibit distinct X-ray emissions thought to be driven by their strong magnetic fields. Here we present the results of magnetar X-ray spectra analysis in their quiescent state. Most of the spectra of magnetars can be fitted with a model consisting of a power-law and a black body component. We found that the luminosity of the power-law component can be described by a function of black-body temperature and emission-region radius. The same relation was seen in pulsars whose X-ray emission mechanism is thought to be different. The fact that magnetars and pulsars share a common fundamental plane in the space spanned by non-thermal X-ray luminosity, surface temperature and the radius of the thermally emitting region presents both challenges and hints to theoretical models for a complete comprehension of the magnetospheric emissions from these two classes of neutron stars.
\end{abstract}




\begingroup
\let\clearpage\relax
\endgroup
\newpage

\section{Introduction} \label{sec:intro}
Magnetars, known as soft gamma repeaters (SGRs) or anomalous X-ray pulsars (AXPs) in observations, exhibit distinctive timing properties and transient X-ray activity. Currently, there are 26 confirmed magnetars \citep{2014ApJS..212....6O}. In contrast to normal pulsars' X-ray emission, which are powered by their rotational energy loss, magnetars derive their X-ray emission primarily from the decay of their internal magnetic fields \citep{1992ApJ...392L...9D,2002ApJ...574..332T}. In the quiescence, X-ray spectra of magnetar ($<$ 10\,keV) are typically described by a combination of a non-thermal power-law and a thermal black body component \citep{2010ApJ...710L.115K,2023JKAS...56...41S}. 

A twisted magnetosphere model has been developed to explain emission mechanism of magnetars \citep{2002ApJ...574..332T}. In this model, the magnetic field lines are anchored to the crust of the neutron star. During the activity of the star, such as crustal motion, the magnetosphere redistributes to a new, stable twisted configuration. Charged particles are then only allowed to move along the closed magnetic field lines in the environment of twisted magnetosphere, forming an optically thick resonant cyclotron scattering region. The thermal photons emitted from the surface of star are Comptonized, leading to a redistribution of their energy and resulting in the observed non-thermal power-law component in the X-ray spectrum. In contrast, the other thermal photons without being scattered continue to travel to the observer, appearing as a black body component in the spectrum.

On the other hand, X-ray emissions from pulsars, except for those from the stellar surface with a thermal nature, are thought to come from certain locations in the magnetosphere or in the wind zone via synchrotron radiation of electron-positron pairs. Recent models for these emissions as well as for gamma rays include the striped-wind model \citep{2015A&A...574A..51P}, the extended-slot-gap and equatorial-current-sheet model \citep{2022ApJ...925..184B}, the synchro-curvature model \citep{2022MNRAS.516.2475I}, and the non-stationary outer-gap model \citep{2017ApJ...834....4T}. These models all propose that the non-thermal X-ray emissions are from a location far away from the neutron star and the mechanism is synchrotron radiation, in contrast to the Comptonization of surface thermal photons in the closed-line regions close to the neutron star for the case of magnetars.

\begin{table*}
\centering
\caption{Timing properties and distance of magnetars.}
\label{tab:mgt_basic}
\renewcommand{\arraystretch}{1.25}
\begin{threeparttable}
\begin{tabularx}{0.87\textwidth}{c|cccccc|cc}
\hline
\hline
Magnetars   & $P$   & $\dot{P}$ & $\tau$ & log $\dot{E}$ & log $B_{\rm{s}}$ & Timing     & Distance  & Distance \\
            & (s)   & (s/s)      &  (yr)     & (erg/s)       &  (G)         & References  & (kpc)     & References \\
\hline
CXOU J010043.1$-$721134 & 8.0204	& 1.88$\times 10^{-11}$     & 6.76$\times 10^{3}$   & 33.2  & 14.6  & 1  & 62.4 $\pm$ 1.6    & 16,17    \\
4U 0142+61              & 8.6885	& 1.94$\times 10^{-12}$     & 7.09$\times 10^{4}$   & 32.1  & 14.1  & 2  & 3.6 $\pm$ 0.4     & 18       \\
SGR 0501+4516           & 5.7628	& 4.58$\times 10^{-12}$     & 1.99$\times 10^{4}$   & 33.0  & 14.2  & 3  & 2.0 $\pm$ 0.4     & 19,20    \\
SGR 0526$-$66           & 8.0565	& 4.02$\times 10^{-11}$     & 3.18$\times 10^{3}$   & 33.5  & 14.8  & 4  & 53.6 $\pm$ 1.2    & 21,22    \\
1E 1048.1$-$5937        & 6.4571	& 3.40$\times 10^{-11}$     & 3.01$\times 10^{3}$   & 33.7  & 14.7  & 5$^{*}$ & 9.0 $\pm$ 1.7     & 18       \\
1E 1547.0$-$5408 (S1)   & 2.0692	& 2.32$\times 10^{-11}$     & 1.42$\times 10^{3}$   & 35.0  & 14.3  & 6$^{*}$ & 4.5 $\pm$ 0.5     & 23       \\
1E 1547.0$-$5408 (S2)   & 2.0902	& 1.58$\times 10^{-11}$     & 2.10$\times 10^{3}$   & 34.8  & 14.3  & 7  & 4.5 $\pm$ 0.5     & 23       \\
CXOU J164710.2$-$455216 & 10.6107	& 9.20$\times 10^{-13}$     & 1.83$\times 10^{5}$   & 31.5  & 14.0  & 8  & 4.06 $\pm$ 0.35   & 24,25    \\
1RXS J170849.0$-$400910 & 11.0023	& 1.95$\times 10^{-11}$     & 8.95$\times 10^{3}$   & 32.8  & 14.7  & 2  & 3.8 $\pm$ 0.5     & 18       \\
CXOU J171405.7$-$381031 & 3.8360	& 4.78$\times 10^{-11}$     & 1.27$\times 10^{3}$   & 34.5  & 14.6  & 9  & 13.2 $\pm$ 0.2    & 26,27,28 \\
SGR 1806$-$20           & 7.7015	& 8.00$\times 10^{-10}$     & 1.53$\times 10^{2}$   & 34.8  & 15.4  & 10 & 8.7 $\pm$ 1.7     & 29,30    \\
XTE J1810$-$197         & 5.5414	& 2.84$\times 10^{-12}$     & 3.09$\times 10^{4}$   & 32.8  & 14.1  & 11 & 2.5 $\pm$ 0.4     & 31       \\
Swift J1822.3$-$1606    & 8.4377	& 2.10$\times 10^{-14}$     & 6.37$\times 10^{6}$   & 30.1  & 13.1  & 12$^{*}$& 1.6 $\pm$ 0.3     & 32       \\
1E 1841$-$045           & 11.7755	& 4.10$\times 10^{-11}$     & 4.56$\times 10^{3}$   & 33.0  & 14.8  & 2  & 5.8 $\pm$ 0.3     & 33,27,34 \\
SGR 1900+14             & 5.2267	& 3.30$\times 10^{-11}$     & 2.51$\times 10^{3}$   & 34.0  & 14.6  & 13 & 12.5 $\pm$ 1.7    & 35,36    \\
SGR 1935+2154           & 3.2474	& 1.43$\times 10^{-11}$     & 3.60$\times 10^{3}$   & 34.2  & 14.3  & 14$^{*}$& 6.6 $\pm$ 0.7     & 37,27,38 \\
1E 2259+586             & 6.9790	& 4.81$\times 10^{-13}$     & 2.30$\times 10^{5}$   & 31.7  & 13.8  & 2  & 3.1 $\pm$ 0.2     & 39,27,40 \\
\hline
PSR J1119$-$6127        & 0.4080	& 4.02$\times 10^{-12}$     & 1.61$\times 10^{3}$   & 36.4  & 13.6  & 15 & 8.4 $\pm$ 0.4     & 41,27,42 \\
\hline
\end{tabularx}
\begin{tablenotes}
\item \textbf{Note.} The different quiescent states of 1E 1547$-$5408 before 2007 and after 2009 outbursts are marked as S1 and S2, respectively. The timing references which do not cover the X-ray data we used are marked with $^{*}$. The magnetar-like pulsar PSR J1119$-$6127 is shown at the bottom of the table. 
\item \textbf{References.} 
(1) \cite{2005ApJ...627L.137M}; (2) \cite{2014ApJ...784...37D}; (3) \cite{2014MNRAS.438.3291C}; (4) \cite{2012MNRAS.424..210G}; (5) \cite{2009ApJ...702..614D}; (6) \cite{2007ApJ...666L..93C}; (7) \cite{2023ApJ...945..153L}; (8) \cite{2007ApJ...664..448I}; (9) \cite{2019ApJ...882..173G}; (10) \cite{2015ApJ...809..165Y}; (11) \cite{2019MNRAS.483.3832P}; (12) \cite{2014ApJ...786...62S}; (13) \cite{2019PASJ...71...90T}; (14) \cite{2016MNRAS.457.3448I}; (15) \cite{2011MNRAS.411.1917W}; (16) \cite{2002ApJ...574L..29L}; (17) \cite{2012AJ....144..107H}; (18) \cite{2006ApJ...650.1070D}; (19) \cite{2010ApJ...722..899G}; (20) \cite{2006Sci...311...54X}; (21) \cite{2004ApJ...609L..13K}; (22) \cite{2012AJ....144..106H}; (23) \cite{2010ApJ...710..227T}; (24) \cite{2006ApJ...636L..41M}; (25) \cite{2022MNRAS.516.1289N}; (26) \cite{2010ApJ...710..941H}; (27) \cite{2022ApJ...940...63R}; (28) \cite{2018MNRAS.477.2243R}; (29) \cite{1997ApJ...478..624C}; (30) \cite{2008MNRAS.386L..23B}; (31) \cite{2020MNRAS.498.3736D}; (32) \cite{2012ApJ...761...66S}; (33) \cite{1997ApJ...486L.129V}; (34) \cite{2018AJ....155..204R}; (35) \cite{2000ApJ...533L..17V}; (36) \cite{2009ApJ...707..844D}; (37) \cite{2014GCN.16533....1G}; (38) \cite{2020ApJ...905...99Z}; (39) \cite{1981Natur.293..202F}; (40) \cite{2018MNRAS.473.1705S}; (41) \cite{2001ApJ...554..152C}; (42) \cite{2004MNRAS.352.1405C} \\
\end{tablenotes}
\end{threeparttable}
\end{table*}

The mechanisms producing non-thermal X-rays from magnetars in quiescence and from pulsars are considered to be very different. However, the identification of two rotation-powered pulsars which displayed magnetar-like burst activity, namely PSR J1119$-$6127 and PSR J1846$-$0258 \cite{2016ApJ...829L..21A,2008Sci...319.1802G}, suggests a potential connection between magnetars and pulsars. In this paper, we explore the relationships between the spectral properties of non-thermal and thermal emissions in magnetars and investigate their similarities to pulsars.

\section{Data} \label{sec:data}
\subsection{Data Selection}
We conducted a search for observations of magnetars in publicly available Chandra or XMM data archives, which offer higher sensitivity and a larger effective area. To study the quiescent state only, we excluded data from the outburst epoch. If the end of the outburst state was not well-defined, we selected data from the epoch with the lowest flux or the greatest temporal separation from the outburst.

The period and period derivative of our samples are chosen at the epoch of the X-ray data we used. If no timing information was available at the epoch of the X-ray data, we derive the period and period derivative using the closest available timing data. Most of the data we used are in the epoch range used for period derivative calculations. The timing properties of magnetars, including the periods $P$ and period derivatives $\dot{P}$, are summarized in Table \ref{tab:mgt_basic}. Other timing variables were derived from the $P$ and $\dot{P}$ values, assuming a moment of inertia of $10^{45}$\,g cm$^{2}$ and a magnetar radius of 10\,km.

The distances of magnetars are mainly taken from the McGill Online Magnetar Catalog \citep{2014ApJS..212....6O} with updated measurement (Table \ref{tab:mgt_basic}). However, the distance to SGR 0501+4516 has not been directly measured, so we adopted a distance of $\sim$ 2\,kpc to the Perseus Arm \citep{2006Sci...311...54X}, with an assumed uncertainty of 40\%, as the distance to SGR 0501+4516.

\begin{table*}
\centering
\caption{Spectral fitting results of magnetars with a PL-plus-BB model.}
\label{tab:spec_PLBB}
\renewcommand{\arraystretch}{1.25}
\begin{threeparttable}
\begin{tabularx}{0.89\textwidth}{cccccccc}
\hline
\hline
Magnetars & $n_{\rm{H}}$            & $\Gamma$ & $kT$  & $R$  & log $L_{\rm{PL}}$  & log $L_{\rm{BB}}$ & $\chi_{\nu}^2$ (d.o.f) \\
          & ($10^{22}$ $cm^{-2}$) &        & (eV)  & (km) & (erg/s) & (erg/s) &                     \\
\hline
CXOU J010043.1$-$721134 & $0.325\substack{+0.064\\-0.082}$	& $3.75\substack{+0.34\\-0.36}$		& $405\substack{+16\\-16}$      & $6.34\substack{+0.69\\-0.59}$     & $35.44\pm0.17$    & $35.14\pm0.04$    & 0.93(254) \\
4U 0142+61              & $0.952\substack{+0.014\\-0.014}$	& $3.692\substack{+0.034\\-0.035}$	& $417.8\substack{+4.5\\-4.3}$  & $6.27\substack{+0.22\\-0.22}$     & $35.65\pm0.10$    & $35.19\pm0.10$    & 0.97(852) \\
SGR 0501+4516           & $1.144\substack{+0.081\\-0.075}$	& $4.50\substack{+0.38\\-0.35}$		& $720\substack{+32\\-28}$      & $0.120\substack{+0.020\\-0.019}$  & $33.94\pm0.22$    & $32.70\pm0.20$    & 0.89(240) \\
SGR 0526$-$66           & $0.45\substack{+0.037\\-0.043}$	& $2.80\substack{+0.11\\-0.14}$		& $434\substack{+23\\-19}$      & $5.29\substack{+0.86\\-0.81}$     & $35.96\pm0.07$    & $35.10\pm0.07$    & 0.94(503) \\
1E 1048.1$-$5937        & $1.08\substack{+0.021\\-0.021}$	& $3.154\substack{+0.056\\-0.055}$	& $575.3\substack{+7.8\\-7.6}$  & $1.863\substack{+0.063\\-0.060}$  & $35.21\pm0.17$    & $34.69\pm0.16$    & 1.02(1317)\\
1E 1547.0$-$5408 (S1)   & $3.20\substack{+0.56\\-0.49}$		& $3.6\substack{+0.7\\-1.1}$		& $397\substack{+31\\-16}$      & $0.84\substack{+0.16\\-0.17}$     & $33.80\pm0.69$    & $33.35\pm0.16$    & 1.13(250) \\
1E 1547.0$-$5408 (S2)   & $3.6\substack{+1.1\\-0.4}$		& $2.1\substack{+1.4\\-2.4}$		& $582\substack{+52\\-50}$      & $1.59\substack{+0.26\\-0.20}$     & $34.25\pm0.89$    & $34.57\pm0.15$    & 1.07(274) \\
CXOU J164710.2$-$455216 & $1.48\substack{+0.49\\-0.37}$		& $2.6\substack{+0.8\\-1.6}$		& $507\substack{+34\\-38}$      & $0.257\substack{+0.027\\-0.070}$  & $32.83\pm0.63$    & $32.75\pm0.16$    & 0.99(166) \\
1RXS J170849.0$-$400910 & $1.454\substack{+0.035\\-0.035}$	& $2.712\substack{+0.071\\-0.076}$	& $446.3\substack{+5.8\\-5.6}$  & $3.06\substack{+0.15\\-0.17}$     & $35.22\pm0.12$    & $34.67\pm0.12$    & 1.03(393) \\
CXOU J171405.7$-$381031 & $3.96\substack{+0.32\\-0.34}$		& $3.14\substack{+0.34\\-0.44}$		& $475\substack{+20\\-22}$      & $2.54\substack{+0.36\\-0.49}$     & $35.30\pm0.29$    & $34.62\pm0.13$    & 0.88(530) \\
SGR 1806$-$20           & $6.12\substack{+0.35\\-0.33}$		& $1.49\substack{+0.12\\-0.14}$		& $635\substack{+75\\-75}$      & $0.97\substack{+0.33\\-0.21}$     & $34.86\pm0.18$    & $34.29\pm0.19$    & 1.04(680) \\
XTE J1810$-$197         & $1.202\substack{+0.073\\-0.062}$  & $7.35\substack{+0.43\\-0.36}$     & $294\substack{+13\\-13}$      & $0.62\substack{+0.15\\-0.14}$     & $34.93\pm0.20$    & $32.63\pm0.20$    & 1.06(1100)\\
Swift J1822.3$-$1606    & $0.77\substack{+0.12\\-0.11}$     & $4.96\substack{+0.39\\-0.40}$     & $91\substack{+13\\-10}$       & $25\substack{+33\\-14}$           & $32.73\pm0.26$    & $33.73\pm0.53$    & 0.91(111) \\
1E 1841$-$045           & $2.42\substack{+0.13\\-0.13}$		& $2.13\substack{+0.25\\-0.28}$		& $426\substack{+21\\-22}$      & $4.78\substack{+0.53\\-0.42}$     & $35.15\pm0.13$    & $34.98\pm0.07$    & 0.95(282) \\
SGR 1900+14             & $2.62\substack{+0.17\\-0.17}$		& $2.55\substack{+0.21\\-0.24}$		& $398\substack{+30\\-33}$      & $4.29\substack{+0.94\\-0.63}$     & $35.17\pm0.18$    & $34.77\pm0.14$    & 1.02(477) \\
SGR 1935+2154           & $2.41\substack{+0.22\\-0.20}$		& $2.45\substack{+0.44\\-0.46}$		& $366\substack{+27\\-29}$      & $1.95\substack{+0.49\\-0.34}$     & $33.84\pm0.26$    & $33.94\pm0.12$    & 0.87(154) \\
1E 2259+586             & $1.117\substack{+0.024\\-0.025}$	& $3.982\substack{+0.073\\-0.074}$	& $414.1\substack{+7.6\\-7.4}$  & $3.12\substack{+0.17\\-0.17}$     & $35.33\pm0.07$    & $34.56\pm0.06$    & 1.03(263) \\
\hline
PSR J1119$-$6127        & $1.04\substack{+0.25\\-0.19}$		& $2.2\substack{+1.0\\-1.1}$		& $299\substack{+42\\-49}$      & $0.68\substack{+0.42\\-0.20}$     & $32.37\pm0.41$    & $32.66\pm0.15$    & 1.02(35)  \\
\hline
\end{tabularx}
\begin{tablenotes}
\item \textbf{Note.} The different quiescent states of 1E 1547$-$5408 before 2007 and after 2009 outbursts are marked as S1 and S2, respectively. The magnetar-like pulsar PSR J1119$-$6127 is shown at the bottom of the table. \\
\end{tablenotes}
\end{threeparttable}
\end{table*}

\subsection{Data Reduction}
The data reduction for Chandra was done by CIAO 4.14 \citep{2006SPIE.6270E..1VF}. We first used the task \verb'chandra_repro' to reproduce the data with the latest calibration information. We excluded any time intervals with magnetar short bursts or high particle background to ensure a good time interval. The source region was defined by a 1$''$-radius circle or 2$''$-long box which enclose approximately 90\% of photons from the target for timed exposure (TE) mode and clock counting (CC) mode, respectively. We used a 4$''$--20$''$ annulus for TE mode and two 16$''$-long boxes on each side of the source for CC mode to define the background region. To extract the magnetar spectrum for further analysis, we performed the task \verb'specextract'.

For XMM data reduction, we used SAS 20.0 \footnote{”XMM-Newton Users Handbook”, Issue 2.20, 2022 (ESA: XMM-Newton SOC).}. Similarly to Chandra data, we began with the tasks \verb'emproc' and \verb'epproc' to reproduce the data with the latest calibration information for MOS and PN camera. Then we selected a good time interval by excluding any time periods with magnetar short bursts or high particle background. We defined the source region with a 20$''$-, 30$''$-, or 40$''$-radius circle, depending on the source's brightness and background level. These regions enclose about 75\% to 90\% of photons from a point source. The background region was chosen to surround the source and on the same detector chip as the source. Lastly, we used task \verb'evselect' to extract the magnetar spectrum.

For a few special cases, we employed different data reduction processes as described below.

For SGR 0526$-$66, located in supernova remnant (SNR) N49 in LMC \citep{1980ApJ...237L...7E}, we aimed to minimize the effect of X-rays from the SNR by selecting only 9 pixels centered on the source coordinate, which corresponds to a 0.75$''$-radius circle enclosing about 80\% of photons from the source. For the background region, we used an ellipse with semi-axes of 2.8$''$ and 2$''$, respectively, surrounding the magnetar while excluding the central 1.2$''$-radius.

As for the magnetar 1E 1841$-$045, located in SNR Kes 73 \citep{1997ApJ...486L.129V}, all the Chandra TE mode observations suffered from pile-up, so the spectral information is unreliable. Therefore, we used a Chandra CC mode observation and chose a smaller, more compact region to better estimate the background photons. We used a 1.8$''$-long box for the source region and two 1.8$''$-long boxes on each side of the source, separated from the source region by one pixel, for the background region.

Following the 2009 outburst of 1E 1547.0$-$5408, it reached a new steady state, which was one order of magnitude brighter than its previous quiescent flux level \citep{2020A&A...633A..31C}. We consider this to be the second quiescent state, and thus we analyzed Swift/XRT data taken in windowed timing mode after 2020. We used the \verb'xrtpipeline' in Heasoft 6.30 \citep{2014ascl.soft08004N} to process the data and excluded high particle background intervals to select good time intervals. The source and background regions were defined by a 50$''$-long box and an 80$''$-long box, respectively, separated by at least 40$''$. Finally, we used the \verb'extractor' task to extract the spectra.

\begin{table*}
\centering
\caption{Average values and standard deviation (in parentheses) of different groups of targets.}
\label{tab:avg_std_PLBB}
\renewcommand{\arraystretch}{1.25}
\begin{threeparttable}
\begin{tabularx}{0.62\textwidth}{ccccccc}
\hline
\hline
Groups &  Number     &  $\Gamma$ & log $L_{\rm{PL}}$ & $kT$ & $R$ & log $L_{\rm{BB}}$  \\
        &          &            & (erg/s) & (eV) & (km) & (erg/s)  \\
\hline
Magnetars       & 15 & 2.97(0.82) & 34.8(0.8) & 480(103) & 2.89(2.09)   & 34.4(0.8)   \\
Pulsars         & 26 & 1.69(0.50) & 31.6(1.5) & 147(93)  & 3.41(6.64)   & 31.9(1.2)   \\
Pulsars(HS)     & 6  & 2.22(0.32) & 29.6(0.4) & 257(48)  & 0.024(0.010) & 29.4(0.5)   \\
XTE J1810       & 1  & 7.35       & 34.9      & 294      & 0.62         & 32.6        \\
Swift J1822     & 1  & 4.96       & 32.7      & 91       & 25           & 33.7        \\
PSR J1119       & 1  & 2.24       & 32.4      & 299      & 0.68         & 32.7        \\
\hline
\end{tabularx}
\begin{tablenotes}
\item \textbf{Note.} The 26 pulsars and 6 pulsars with hot spot (HS) are those studied in  \citet{2023MNRAS.520.4068C}. \\
\end{tablenotes}
\end{threeparttable}
\end{table*}

\subsection{Spectral Fitting}
After the data reduction, XSPEC 12.12 of Heasoft 6.30 \citep{2014ascl.soft08004N} was then used for spectral fitting. We selected an energy range of 0.5--7\,keV and 0.3--8\,keV for Chandra and XMM, respectively. To ensure better statistics, we required each bin to have at least 20 photons when grouping the source spectra. We fitted each magnetar spectrum using a power-law (PL) plus black body (BB) model. A constant was added to the model to account for the normalization of different observations when joint-fitting multiple spectra of the same magnetar. The fitting of the X-ray data selected for the same source yielded consistent results. We considered a fit to be acceptable only if the p-value of $\chi^2$ distribution was larger than 0.05. The fitting results are shown in the Table \ref{tab:spec_PLBB}. In order to compare with previous work, the black body and power-law luminosities were calculated in the range of 0--100\,keV to represent bolometric luminosity and 0.5--8\,keV, respectively. All the uncertainties reported in this paper are 1$\sigma$ errors.

For SGR 0418+5729, SGR 1627$-$41, SGR J1745$-$2900, and Swift J1834.9$-$0846, their quiescent fluxes are dimmer compared to other magnetars, resulting in larger errors in the extracted spectra. Fitting these spectra with a single component model (PL or BB) yielded $\chi_{\nu}^2$ of 0.58--0.75, indicating that the data were not of sufficient quality to fit with multiple components. Therefore, these sources were excluded from our sample. For Swift J1555.2$-$5402, PSR J1622$-$4950, Swift J1818.0$-$1607, SGR 1830$-$0645, SGR 1833$-$0832, and 3XMM J185246.6+003317, the quiescent fluxes were too dim to be detected, and hence no spectral information was available for these magnetars. Apart from the above cases, all other magnetars can be fitted with a PL-plus-BB model (Table \ref{tab:spec_PLBB}).

Two pulsars with magnetar-like bursts, PSR J1119$-$6127 and PSR J1846$-$0258 \citep{2016ApJ...829L..21A,2008Sci...319.1802G}, were also considered in this study. However, the Chandra observations of PSR J1846$-$0258, located in the SNR Kes 75 \citep{2000ApJ...542L..37G}, were piled-up so that spectral analysis is unreliable. Hence, only the data of PSR J1119$-$6127 were included for further analysis.

\section{Results}
\subsection{Linear Correlation Coefficient}
After the spectral fitting, we have 17 magnetar samples and a magnetar-like pulsar which can be fitted with a PL-plus-BB model. On average, magnetars have similar black body emitting radius $R$ to pulsars, but higher temperature $T$, resulting in brighter black body luminosity $L_{\rm{BB}}$. Additionally, the power-law index $\Gamma$ and luminosity $L_{\rm{PL}}$ for magnetars are larger than that of pulsars \citep{2010ApJ...710L.115K,2023JKAS...56...41S,2021MNRAS.502..390H,2023MNRAS.520.4068C}.In \citet{2023MNRAS.520.4068C}, 32 pulsar samples were fitted with a PL-plus-BB model. Comparison to the average values and the standard deviation of spectral parameters of magnetars in this work and pulsars studied by \citet{2023MNRAS.520.4068C} are listed in Table \ref{tab:avg_std_PLBB}. The pulsar with magnetar-like burst, PSR J1119$-$6127 (PSR J1119) \citep{2016ApJ...829L..21A,2016ApJ...829L..25G}, is clearly not in the group of magnetars. Hence, PSR J1119$-$6127 was excluded in the following analysis. Moreover, two magnetars, XTE J1810$-$197 (XTE J1810) and Swift J1822.3$-$1606 (Swift J1822), exhibited exceptional characteristics as outliers. XTE J1810$-$197 displayed an unusually large power index of 7.35, which is considered unphysical. Swift J1822.3$-$1606 exhibited a remarkably low black body temperature and an large black body emitting radius of 25\,km. Consequently, these two sources were excluded from magnetar samples.

Firstly, we examined the linear correlation coefficient between the spectral and timing parameters. These spectral parameters include $\Gamma$, $L_{\rm{PL}}$, $T$, $R$, and $L_{\rm{BB}}$. The timing parameters are period $P$, period time-derivative $\dot{P}$, as well as derived quantities from $P$ and $\dot{P}$ such as frequency time-derivative $\dot{\nu}$, characteristic age $\tau$, spin-down power $\dot{E}$, the dipole magnetic field strength at the stellar surface $B_{\rm{s}}$ and at the light cylinder $B_{\rm{lc}}$. We found a moderately strong relationship between $\Gamma$ and $\dot{\nu}$ or $B_{\rm{s}}$, which is consistent with similar study previously \citep{2010ApJ...710L.115K,2023JKAS...56...41S}. Additionally, other timing parameters, such as $\dot{P}$ and $\tau$, displayed similar levels of correlation with $\Gamma$ \citep{2023JKAS...56...41S}. The absolute Pearson correlation coefficients of these four relations ranged from 0.46 to 0.59. On the other hand, none of the other spectral parameters appeared to have a strong relationship with the timing parameters with Pearson correlation coefficient ranging from 0.01 to 0.36.

\begin{figure*}
\includegraphics[width=\textwidth]{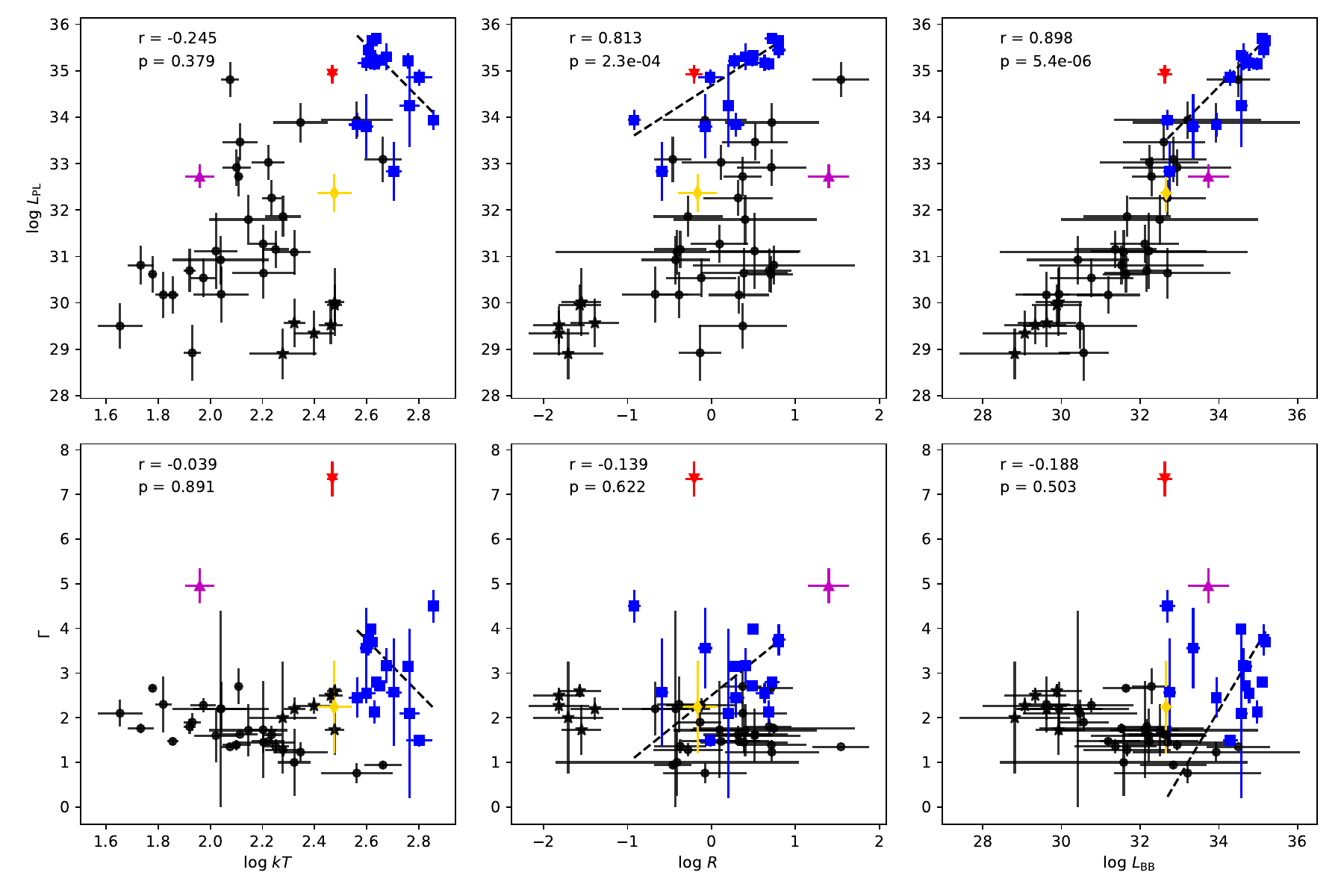}
\caption{Non-thermal spectral properties versus thermal spectral properties. Units are same as those presented in Table  \ref{tab:spec_PLBB}. The blue squares are magnetars. The red and magenta triangles are XTE J1810$-$197 and Swift J1822.3$-$1606, respectively. The yellow diamond is PSR J1119$-$6127. The black circles and stars are the 26 pulsars and 6 pulsars with hot spot taken from \citet{2023MNRAS.520.4068C}. The Pearson correlation coefficients (r) and their corresponding p-values for the magnetar samples are shown in the legend. The dashed lines are the best linear fits for the magnetar samples.}
\label{fig:PLvsBB}
\end{figure*}

The linear correlation coefficient between different spectral parameters was also examined. We found a strong correlation between the non-thermal power-law luminosity $L_{\rm{PL}}$ and the thermal spectral parameters, specifically the black body luminosity $L_{\rm{BB}}$ and emitting radius $R$. The Pearson correlation coefficients were 0.898 and 0.813, respectively, with corresponding random probabilities of 5.4$\times 10^{-6}$ and 2.3$\times 10^{-4}$, indicating a high level of statistical significance. We recalculated the correlation coefficients using power-law flux, black body flux, and the normalization of the black body, which are all obtained purely from the spectral fitting and are not influenced by the distance factor. The resulting correlation coefficients remained similar, with values of 0.903 and 0.783. The non-thermal spectral properties versus thermal spectral properties plots are shown in Figure \ref{fig:PLvsBB}, including the pulsar sample studied by \citet{2023MNRAS.520.4068C} for comparison. In general, magnetars and pulsars do not form two distinct groups except that magnetars are with higher luminosity and temperature. The weak anti-correlation between $L_{\rm{PL}}$ and $kT$ in the magnetar sample should not be taken seriously. A more obvious positive correlation shows up when magnetars and pulsars are jointly considered. From these plots, it is evident that XTE J1810$-$197 and Swift J1822.3$-$1606 behave differently from other magnetars. 

\subsection{Parameters Fitting}
In light of the strong correlation observed between the non-thermal and thermal spectral parameters, we try to fit one spectral parameter ($\Gamma$, $L_{\rm{PL}}$, $T$, $R$, $L_{\rm{BB}}$) by two of other spectral ($\Gamma$, $T$, $R$) and timing parameters ($P$, $\dot{P}$). Most of the combinations resulted in poor fits with $\chi_{\nu}^2$ ranging from 2.95 to 62.24. Remarkably, we found that the $L_{\rm{PL}}$ as a function of $T$ and $R$ yielded a $\chi_{\nu}^2$ of 1.455. The p-value corresponding to this fitting is 0.133, which is not low enough to reject the null hypothesis. We therefore considered this to be a good fit. When considering $T$ or $R$ alone in the fitting, the $\chi_{\nu}^2$ were noticeably higher, with values of 6.39 and 2.80, respectively. The application of an F-test clearly indicates the necessity and improvement achieved by incorporating both $T$ and $R$ in the $L_{\rm{PL}}$ versus $T$ and $R$ fitting. 

\begin{figure*}
\includegraphics[width=\textwidth]{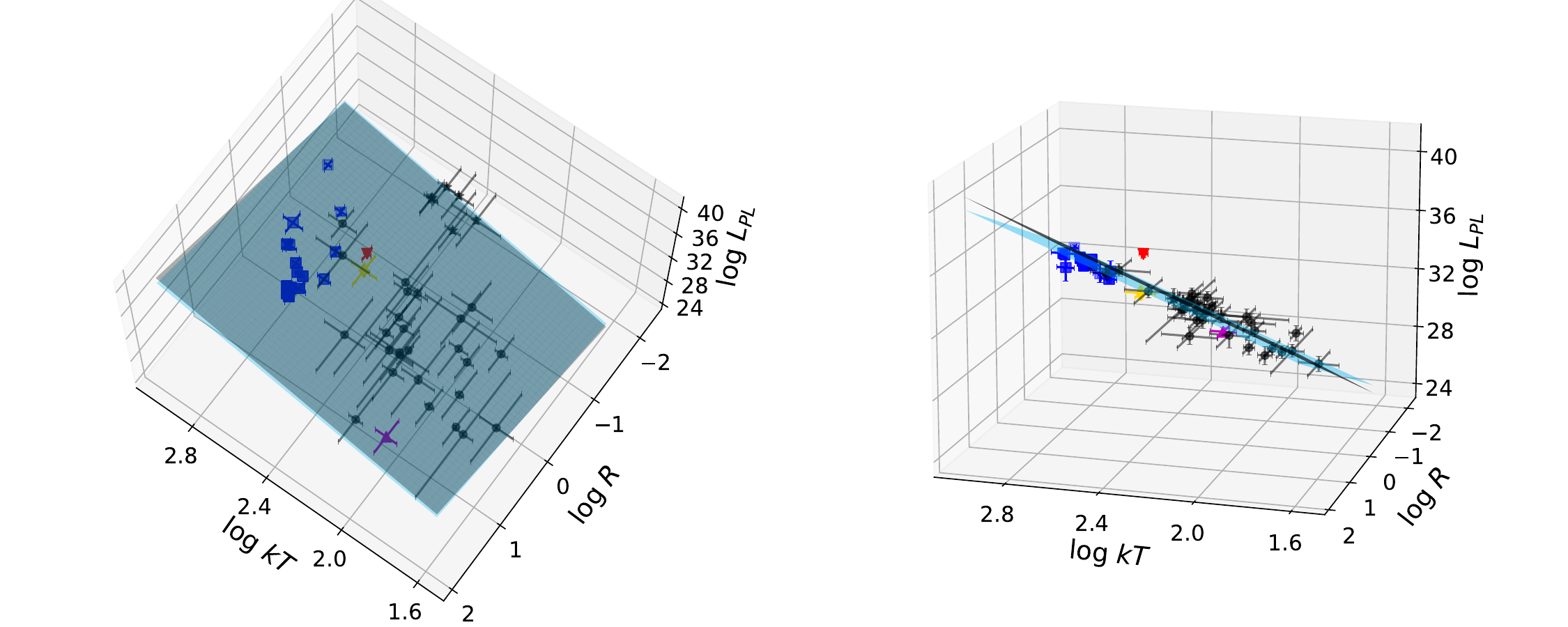}
\caption{The fundamental planes in the space of power-law luminosity $L_{\rm{PL}}$ (erg/s), black body temperature $kT$ (eV), and black body emitting radius $R$ (km). The data points are same as those in Figure \ref{fig:PLvsBB}. The blue and gray planes are the $L_{\rm{PL}}$-$kT$-$R$ planes of magnetars and pulsars, respectively. The left panel is a more face-on view to show the different neutron star groups. The right panel is a more edge-on view to show the planes.}
\label{fig:LPL_TR}
\end{figure*}

Moreover, we found that magnetars and pulsars share the common $L_{\rm{PL}}$-$T$-$R$ plane \citep{2023MNRAS.520.4068C}, despite the differences in the distribution of spectral parameters between the two groups (Table \ref{tab:avg_std_PLBB}). The $L_{\rm{PL}}$-$T$-$R$ planes of magnetars and pulsars are presented in Figure \ref{fig:LPL_TR}. Furthermore, the magnetar-like pulsar PSR J1119$-$6127 and the outlier magnetar Swift J1822.3$-$1606 both lie on this plane. When including these two neutron stars in the magnetar samples, the fitting of $L_{\rm{PL}}$-$T$-$R$ improved, reducing the $\chi_{\nu}^2$ from 1.455 to 1.282. However, if we include another outlier XTE J1810$-$197 in the magnetar sample, the fitting becomes unacceptable with a significantly higher $\chi_{\nu}^2$ of 3.906, further suggesting that its spectral description is unphysical. Subsequently, we performed a fit including all available neutron star samples, including magnetars, pulsars, PSR J1119$-$6127, and Swift J1822.3$-$1606. The resulting function is presented in Table \ref{tab:LPL_TR}. 

\section{Discussion}
\subsection{The Common Fundamental Plane}
The non-thermal power-law luminosity $L_{\rm{PL}}$ of our neutron star samples can be described by their black body temperature $T$ and emitting radius $R$ approximately as 
\begin{equation}
L_{\rm{PL}} \propto T^{5.5}R^{2} ,
\end{equation}
or equivalently, 
\begin{equation}
L_{\rm{PL}} \propto T^{1.5}L_{\rm{BB}} .
\end{equation}
This remarkable relationship suggests that the thermal black body radiation emitted by these neutron stars plays a crucial role in determining their non-thermal power-law luminosity. Since both $T$ and $L_{\rm{BB}}$ play significant roles in determining $L_{\rm{PL}}$, it is likely that the process of generating non-thermal photons does not occur in a region close to the stellar surface. Otherwise, only $T$ would be crucial in this process.

In \citet{2023MNRAS.520.4068C}, it was reported that, for pulsars, the power index $\Gamma$ can be describe relatively well by $T$ and $R$ with a fitting $\chi_{\nu}^2$ at 1.34. However, when applying the same fitting to magnetar samples, the results were considerably poorer, with a $\chi_{\nu}^2$ value of 13.55. The $\Gamma$-$T$-$R$ plane for pulsars is not observed in magnetars. 

As introduced in the Section \ref{sec:intro}, the non-thermal X-ray emission mechanisms of magnetars and pulsars are thought to differ significantly. The non-thermal X-ray emission from magnetars is the Comptonization of surface thermal photons. Pulsars emit non-thermal X-rays from synchrotron radiation of electron-positron pairs. The common relation between thermal and non-thermal emissions of magnetars and pulsars intrigues a more detailed examination of our current understanding.

\begin{table}
\centering
\caption{Fitting results of log $L_{\rm{PL}}$ = $\alpha$ log $kT$ + $\beta$ log $R$ + $c$ of different target groups.}
\label{tab:LPL_TR}
\renewcommand{\arraystretch}{1.25}
\begin{threeparttable}
\begin{tabularx}{0.47\textwidth}{cccccc}
\hline
\hline
Groups &  $\alpha$  &  $\beta$  & c & $\chi_{\nu}^2$ (d.o.f) \\
\hline
Magnetars               & $5.79\pm1.87$ & $1.94\pm0.32$ & $19.0\pm5.1$ & 1.455(12)   \\
Magnetars$\dagger$      & $6.38\pm1.02$ & $2.04\pm0.24$ & $17.3\pm2.8$ & 1.282(14)   \\
Pulsars                 & $5.96\pm0.64$ & $2.24\pm0.18$ & $18.7\pm1.4$ & 0.490(29)   \\
All                     & $5.67\pm0.25$ & $2.02\pm0.09$ & $19.2\pm0.6$ & 0.746(46)   \\
\hline
\end{tabularx}
\begin{tablenotes}
\item \textbf{Note.} $\dagger$ Swift J1822.3$-$1606 and PSR J1119$-$6127 are included in this magnetar group fitting.\\
\end{tablenotes}
\end{threeparttable}
\end{table}

Several models of magnetars based on the twisted magnetosphere model have been proposed to explain various unique phenomena observed in magnetars successfully. For instance, these models can account for the hard X-ray component ($>$20\,keV) in quiescence \citep{2005ApJ...634..565T,2007ApJ...657..967B,2013ApJ...762...13B}, bursting activities \citep{2009ApJ...703.1044B,2023MNRAS.519.4094W}, and variable radio emission \citep{2008ApJ...688..499T,2019ApJ...875...84W}. In contrast, non-thermal X-ray emission from pulsars, which is quite steady, is considered together with emissions in gamma-rays or optical bands as originated from electron-positron pairs somewhere in the dipolar co-rotating magnetosphere or in the wind zone \citep{2015A&A...574A..51P,2017ApJ...834....4T,2022ApJ...925..184B,2022MNRAS.516.2475I}. Despite their distinct differences, these models all imply some link between their non-thermal X-ray emission and surface thermal emission.

In twisted magnetosphere magnetar models, persistent non-thermal X-rays come from the Comptonization of surface thermal photons in closed field-line regions close to the stellar surface \citep{2002ApJ...574..332T}. The particles performing Comptonization can be electron-positron pairs generated through one-photon pair production. Accelerated seed particles in the closed field line up-scatter thermal X-ray photon to the energy $>$ MeV. These high-energy photons, interacting with the strong magnetic field, subsequently generate pairs that are accelerated to scatter more photons, initiating a pair-production avalanche. Consequently, a Comptonization region for thermal photon forms within the closed field line. The existence of a relation between thermal and non-thermal X-rays seems obvious. The energy distribution of those pairs certainly play a role in that relation. 

In pulsar's outer-gap models, the electron-positron pairs responsible for non-thermal X-ray emission are created through two-photon pair production. High-energy GeV photons emitted by charged particles accelerated in the outer gap collide with surface thermal photons to make pairs. These pairs may also be accelerated to emit GeV photons to make more pairs. A pair-production avalanche is developed so that the space charge density can meet the Goldreich-Julian condition, and the gap region stops growing. X-ray emission comes from the synchrotron radiation of these pairs. Apparently, the energy distribution of these pairs is related to the surface thermal emission. Hence comes the relation between thermal and non-thermal X-rays. On the other hand, in slot-gap or wind-zone models, pair plasma responsible for the non-thermal X-ray synchrotron radiation comes from pair-production avalanche above polar-cap regions, in particular the region close to the last open field lines. This avalanche is activated by the electric acceleration along the magnetic field lines and one-photon pair production in strong magnetic fields in the polar-gap region. It is not obvious how surface thermal photons can play a role in determining non-thermal X-ray luminosities. One should note, however, the energetic of the electrically accelerated charged particles, which emit high-energy photons to make pairs in the strong magnetic field  above the polar cap, may be affected by the inverse Compton scattering between them and surface thermal photons \citep{1995A&A...301..456C,1995ApJ...446..292S,1998ApJ...508..328H,2002ApJ...568..862H,2019ApJ...871...12T}. This provides a possible link between thermal and non-thermal X-ray emissions.

\begin{figure}
\includegraphics[width=\columnwidth]{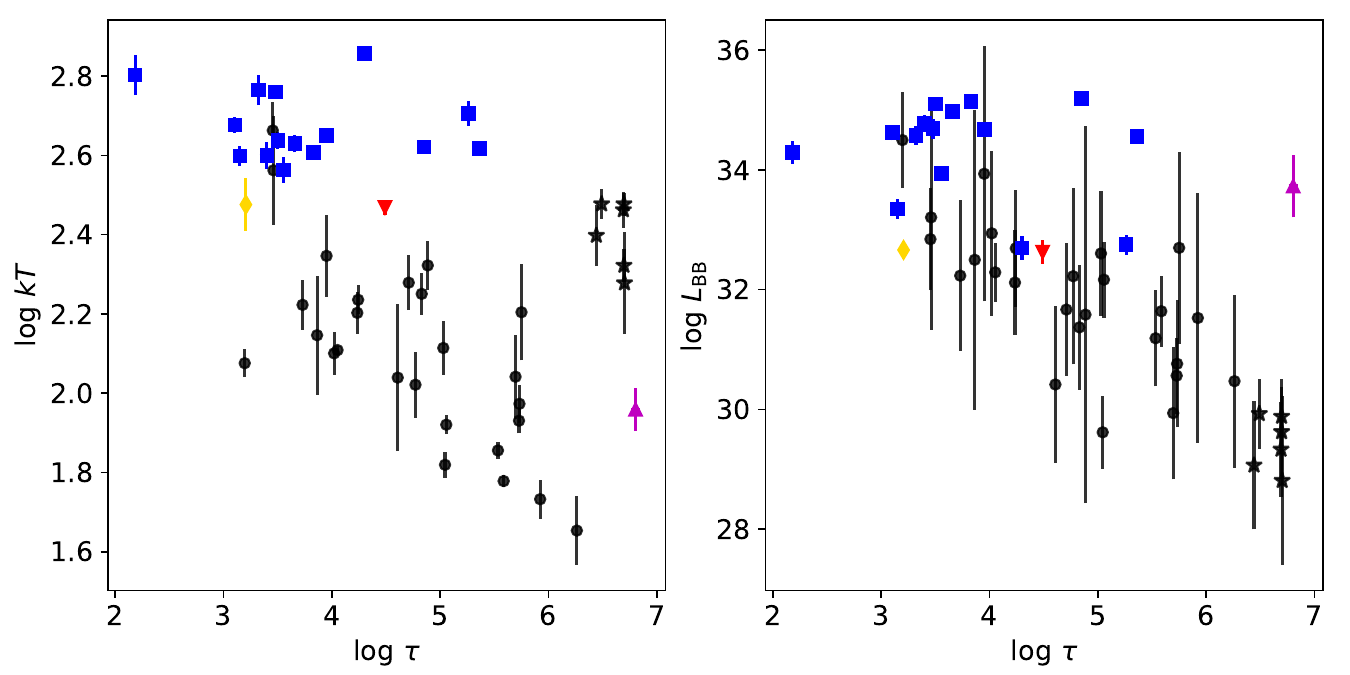}
\caption{Black body temperature (left) and luminosity (right) versus characteristic age. Units are same as those presented in Table \ref{tab:mgt_basic} and \ref{tab:spec_PLBB}. The data points are same as those in Figure \ref{fig:PLvsBB}.}
\label{fig:cooling}
\end{figure}

Models for magnetars and pulsars provide an explanation for the connection between thermal and non-thermal emissions. These models emphasize the role of particles responsible for emitting or scattering photons, as they significantly contribute to the overall emission mechanism. The energy distribution of these particles is strongly influenced by magnetic fields or electric potentials, which can be described using timing parameters. The fundamental plane that determines non-thermal luminosity solely based on thermal emission is then intriguing. However, we note that thermal emission itself can also be described by timing parameters to some extent. This is related to neutron star cooling. In Figure \ref{fig:cooling}, we show the plots of $T$ and $L_{\rm{BB}}$ versus $\tau$. Both $T$ and $L_{\rm{BB}}$ display a clear anti-correlation with $\tau$, but with significant scatter. This scattered relation suggests that the cooling process of neutron stars is not solely determined by their age, but is also influenced by other factors, such as the magnetic field \citep{2009A&A...496..207P} and many others \citep[e.g.][]{2020MNRAS.496.5052P}. As a result, we do not observe tight relationships between $T$/$L_{\rm{BB}}$ and $\tau$. Moreover, magnetars exhibit a larger scatter in these relations, which could be attributed to magnetic reheating during their X-ray outbursts \citep{2016ApJ...833..261B}.

\begin{table*}
\centering
\caption{Fitting results of log $L_{\rm{PL}}$ = $\alpha$ log $kT$ + $\beta$ log $R$ + $\gamma$ log $\dot{P}$ + $c$ of different target groups.}
\label{tab:LPL_TRP}
\renewcommand{\arraystretch}{1.25}
\begin{threeparttable}
\begin{tabularx}{0.59\textwidth}{ccccccc}
\hline
\hline
Groups &  $\alpha$  &  $\beta$  & $\gamma$ & c & $\chi_{\nu}^2$ (d.o.f) \\
\hline
Magnetars           & $7.00\pm1.54$ & $2.04\pm0.25$ & $-0.22\pm0.07$ & $13.3\pm4.6$ & 0.760(11) \\
Magnetars$\dagger$  & $7.22\pm0.83$ & $2.08\pm0.18$ & $-0.22\pm0.06$ & $12.6\pm2.7$ & 0.655(13) \\
Pulsars             & $5.58\pm0.79$ & $2.08\pm0.28$ & $-0.17\pm0.23$ & $21.8\pm4.4$ & 0.499(28) \\
All                 & $6.27\pm0.31$ & $2.11\pm0.09$ & $-0.17\pm0.05$ & $15.7\pm1.3$ & 0.618(45) \\
\hline
\end{tabularx}
\begin{tablenotes}
\item \textbf{Note.} $\dagger$ Swift J1822.3$-$1606 and PSR J1119$-$6127 are included in magnetar group fitting.\\
\end{tablenotes}
\end{threeparttable}
\end{table*}

\begin{table*}
\centering
\caption{Fitting results of log $L_{\rm{PL}}$ = $\alpha$ log $kT$ + $\beta$ log $R$ + $\gamma$ log $\dot{P}$ + $\delta$ log $P$ + $c$ of different target groups.}
\label{tab:LPL_TRPP}
\renewcommand{\arraystretch}{1.25}
\begin{threeparttable}
\begin{tabularx}{0.7\textwidth}{cccccccc}
\hline
\hline
Groups &  $\alpha$  &  $\beta$  & $\gamma$ & $\delta$ & c & $\chi_{\nu}^2$ (d.o.f) \\
\hline
Magnetars           & $7.02\pm1.65$ & $2.04\pm0.28$ & $-0.22\pm0.07$ & $-0.03\pm0.44$ & $13.2\pm4.9$ & 0.836(10) \\
Magnetars$\dagger$  & $7.17\pm0.91$ & $2.06\pm0.21$ & $-0.22\pm0.06$ & $ 0.06\pm0.34$ & $12.7\pm2.8$ & 0.708(12) \\
Pulsars             & $6.17\pm1.14$ & $2.38\pm0.48$ & $ 0.09\pm0.27$ & $ 0.48\pm0.59$ & $19.9\pm5.4$ & 0.505(27) \\
All                 & $6.56\pm0.45$ & $2.17\pm0.11$ & $-0.17\pm0.05$ & $-0.14\pm0.15$ & $15.0\pm1.5$ & 0.619(44) \\
\hline
\end{tabularx}
\begin{tablenotes}
\item \textbf{Note.} $\dagger$ Swift J1822.3$-$1606 and PSR J1119$-$6127 are included in magnetar group fitting.\\
\end{tablenotes}
\end{threeparttable}
\end{table*}

\subsection{Possible higher dimensional fundamental plane}
To explore the possible role of timing properties, we include $P$ or $\dot{P}$ to the $L_{\rm{PL}}$-$T$-$R$ fitting for the magnetar sample. The inclusion of $P$ does not lead to an improvement in the fitting, resulting in a $\chi_{\nu}^2$ of 1.565. The negligible power of $P$ in this fitting suggests that it does not play a significant role. On the other hand, when $\dot{P}$ is included, the $\chi_{\nu}^2$ becomes significantly smaller, from 1.455 to 0.760. The application of the F-test suggested that the addition of the $\dot{P}$ is reasonable. The moderately strong relationship between $\Gamma$ and other $\dot{P}$-related timing properties, as found in this study and previous works \citep{2010ApJ...710L.115K,2023JKAS...56...41S}, further emphasizes the role of $\dot{P}$. When considering only the pulsar sample in the $L_{\rm{PL}}$-$T$-$R$-$\dot{P}$ fitting, the power of $\dot{P}$ is not well constrained due to the very small $\chi_{\nu}^2$ obtained from the $L_{\rm{PL}}$-$T$-$R$ fitting. However, when all the samples are included, a similar relationship is obtained from the fitting. The values of the best-fit parameters are consistent across all the fittings with different sample groups. The detailed fitting results of $L_{\rm{PL}}$ versus $T$, $R$, and $\dot{P}$ for different sample groups are presented in Table \ref{tab:LPL_TRP}. 

Although $P$ does not seem to  play a role, as described above,  to facilitate a full range of discussion on the timing-inferred properties, $P$ and $\dot{P}$ are both included to the $L_{\rm{PL}}$-$T$-$R$ fitting, whose results are listed in Table \ref{tab:LPL_TRPP}. Similar to the previous fittings, the inclusion of $P$ does not lead to an improved fitting, as evident from the obtained $\chi_{\nu}^2$ values and the almost negligible value of the parameter $\delta$ associated with $P$. We also investigated the existence of a $\Gamma$ fundamental plane for magnetars, similar to that observed for pulsars. After incorporating $P$ and $\dot{P}$ in the $\Gamma$-$T$-$R$ fitting for pulsars, the $\chi_{\nu}^2$ decreased from 1.34 to 0.84. The resulting function from this fitting is given as $\Gamma = -5.8\log T -2.3\log R -1.2\log P +0.9\log \dot{P} + constant$ \citep{2023MNRAS.520.4068C}. Similarly, the inclusion of $P$ and $\dot{P}$ improved the $\Gamma$-$T$-$R$ fitting for magnetars, reducing the $\chi_{\nu}^2$ from 13.55 to 3.44. However, it is still an unacceptable fit and the resulting function remained significantly different from that observed for pulsars, with $\Gamma = 3.4\log T +0.6\log R -1.1\log P -0.8\log \dot{P} + constant$. Once again, the fundamental plane that determines $\Gamma$ in pulsars is not observed in magnetars.

\begin{table*}
\centering
\caption{Fitting results of log $L_{\rm{PL}}$ = $\alpha$ log $kT$ + $\beta$ log $R$ + $\gamma$ log $\dot{E}$ + $\delta$ log $B_{\rm{s}}$ + $c$ of different target groups.}
\label{tab:LPL_TREBs}
\renewcommand{\arraystretch}{1.25}
\begin{threeparttable}
\begin{tabularx}{0.7\textwidth}{cccccccc}
\hline
\hline
Groups &  $\alpha$  &  $\beta$  & $\gamma$ & $\delta$ & c & $\chi_{\nu}^2$ (d.o.f) \\
\hline
Magnetars           & $7.02\pm1.65$ & $2.04\pm0.28$ & $-0.05\pm0.11$ & $-0.34\pm0.24$ & $22.0\pm4.1$ & 0.836(10) \\
Magnetars$\dagger$  & $7.17\pm0.91$ & $2.06\pm0.21$ & $-0.07\pm0.09$ & $-0.30\pm0.20$ & $21.9\pm2.5$ & 0.708(12) \\
Pulsars             & $6.17\pm1.13$ & $2.38\pm0.48$ & $-0.10\pm0.18$ & $ 0.38\pm0.42$ & $16.9\pm6.5$ & 0.505(27) \\
All                 & $6.56\pm0.45$ & $2.17\pm0.11$ & $-0.007\pm0.04$ & $-0.32\pm0.11$ & $21.6\pm1.6$ & 0.619(44) \\
\hline
\end{tabularx}
\begin{tablenotes}
\item \textbf{Note.} $\dagger$ Swift J1822.3$-$1606 and PSR J1119$-$6127 are included in magnetar group fitting.\\
\end{tablenotes}
\end{threeparttable}
\end{table*}

\begin{table*}
\centering
\caption{Fitting results of log $L_{\rm{PL}}$ = $\alpha$ log $kT$ + $\beta$ log $R$ + $\gamma$ log $\dot{E}$ + $\delta$ log $B_{\rm{lc}}$ + $c$ of different target groups.}
\label{tab:LPL_TREB}
\renewcommand{\arraystretch}{1.25}
\begin{threeparttable}
\begin{tabularx}{0.7\textwidth}{cccccccc}
\hline
\hline
Groups &  $\alpha$  &  $\beta$  & $\gamma$ & $\delta$ & c & $\chi_{\nu}^2$ (d.o.f) \\
\hline
Magnetars           & $7.02\pm1.65$ & $2.04\pm0.28$ & $-0.56\pm0.28$ & $ 0.68\pm0.48$ & $33.4\pm7.8$ & 0.833(10) \\
Magnetars$\dagger$  & $7.17\pm0.91$ & $2.06\pm0.21$ & $-0.53\pm0.24$ & $ 0.61\pm0.40$ & $32.0\pm6.5$ & 0.706(12) \\
Pulsars             & $6.17\pm1.13$ & $2.38\pm0.48$ & $ 0.48\pm0.64$ & $-0.77\pm0.83$ & $4.2\pm18.9$ & 0.504(27) \\
All                 & $6.57\pm0.45$ & $2.17\pm0.11$ & $-0.49\pm0.16$ & $ 0.65\pm0.23$ & $32.5\pm4.2$ & 0.619(44) \\
\hline
\end{tabularx}
\begin{tablenotes}
\item \textbf{Note.} $\dagger$ Swift J1822.3$-$1606 and PSR J1119$-$6127 are included in magnetar group fitting.\\
\end{tablenotes}
\end{threeparttable}
\end{table*}

A certain combination of $P$ and $\dot{P}$ can always be expressed in terms of another combination of two timing-inferred physical parameters, such as the spin-down power $\dot{E}$ and dipole magnetic field strength at the stellar surface $B_{\rm{s}}$ or at the light cylinder $B_{\rm{lc}}$, which may be more physically relevant to the non-thermal emission. One may directly convert the results in Table \ref{tab:LPL_TRPP} into relationships among $L_{\rm{PL}}$, $T$, $R$, $\dot{E}$, and $B_{\rm{s}}$ or $B_{\rm{lc}}$. One may also conduct fittings among these quantities, whose results are shown in Table \ref{tab:LPL_TREBs} and \ref{tab:LPL_TREB}, which are consistent with those converted from Table \ref{tab:LPL_TRPP}. Based on Tables \ref{tab:LPL_TREBs} and \ref{tab:LPL_TREB}, the power-law luminosity can be described by the black body temperature, black body luminosity, spin-down power, and magnetic field strength as
\begin{equation}
L_{\rm{PL}} \propto T^{2.5}L_{\rm{BB}}^{1.0}\dot{E}^{0.0}B_{\rm{s}}^{-0.3} ,
\end{equation} 
or alternatively,
\begin{equation}
\label{eq:LTREB}
L_{\rm{PL}} \propto T^{2.5}L_{\rm{BB}}^{1.0}\dot{E}^{-0.5}B_{\rm{lc}}^{0.6} .
\end{equation}
The dependence on timing parameters in these relationships is weak compared to that on thermal properties.

\begin{figure*}
\includegraphics[width=\textwidth]{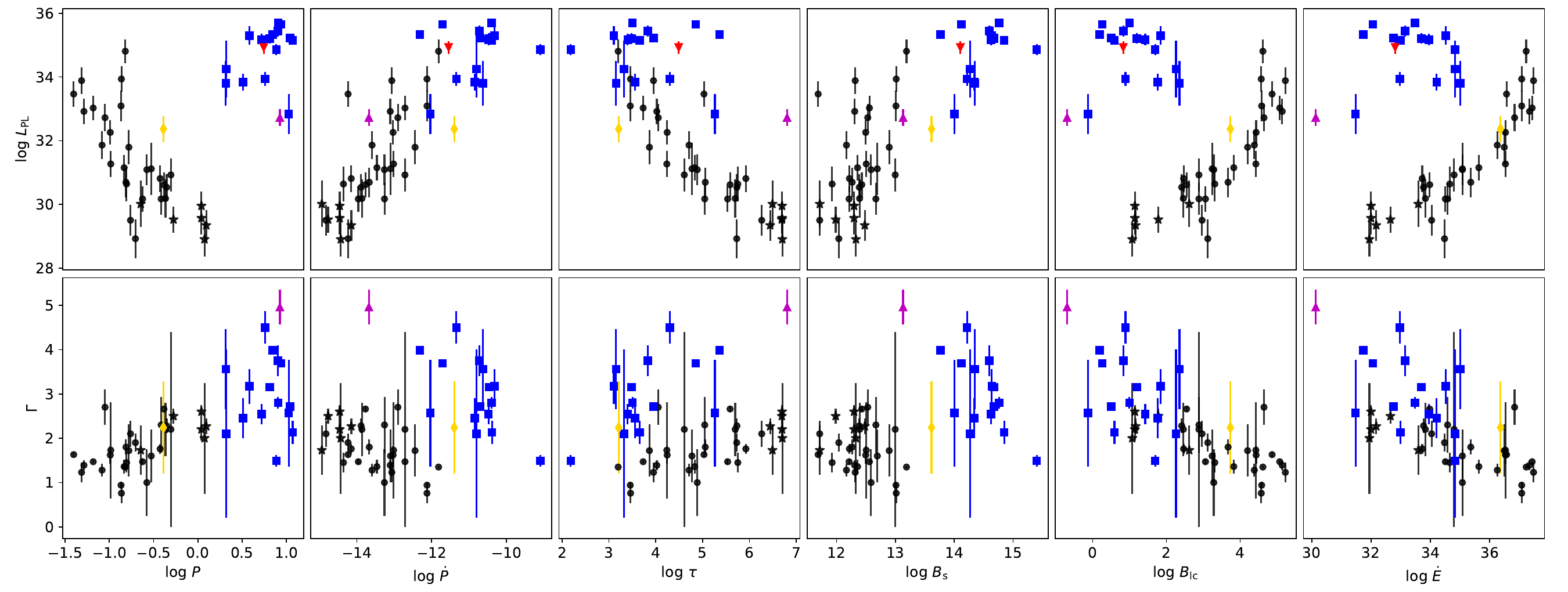}
\caption{Power-law luminosity (upper panels) and index (lower panels) versus timing properties. Units are same as those presented in Table \ref{tab:mgt_basic} and \ref{tab:spec_PLBB}. The data points are same as those in Figure \ref{fig:PLvsBB}. XTE J1810$-$197, the magentar with unphysical $\Gamma$ of 7.35, are not plotted in lower panels for clarity.}
\label{fig:spvst}
\end{figure*}

Although implications of the connection between non-thermal emission, thermal emission, and other key physical parameters can be found in almost all models for magnetars and pulsars, the origins of these models actually differ a lot. For pulsars, the fundamental plane reported here may offer one more diagnostic tool to distinguish different models. The reason why magnetars and pulsars share a common fundamental plane remains puzzling. One possibility, although not really likely, is that the pulsar models could be applicable to magnetars, or vice versa. However, it is important to note that magnetars and pulsars exhibit distinct characteristics and behaviors in the relationships between spectral and timing properties, as shown in Figure \ref{fig:spvst}. Therefore, it is more conceivable that there may be crucial elements missing in these models, which could explain this observed commonality.

In Figure \ref{fig:spvst}, we present plots of non-thermal spectral properties versus timing properties for magnetars and pulsars. Notably, magnetars exhibit significantly different relationships compared to pulsars in terms of $L_{\rm{PL}}$ versus $\dot{E}$ or $B_{\rm{lc}}$. Therefore, it appears more appropriate to consider $B_{\rm{s}}$, which displays a consistent trend for both magnetars and pulsars in the $L_{\rm{PL}}$ versus $B_{\rm{s}}$ relationship, to explain the role of timing parameters in the common fundamental plane. However, the strong relationships observed in $L_{\rm{PL}}$ versus $\dot{E}$ and versus $B_{\rm{lc}}$ for the pulsar group should not be overlooked \citep{2023MNRAS.520.4068C}. It is important to note that these strong relationships may become less prominent when additional parameters are incorporated into a higher-dimensional plane, which may have different physical interpretations. In this higher-dimensional space, we observe a common fundamental plane shared by magnetars and pulsars. This plane represents the missing connection between these two types of objects, while their distinct properties cause their separation on the plane. Furthermore, certain unique properties specific to pulsars contribute to the stronger correlations observed between $L_{\rm{PL}}$ and $\dot{E}$, which are not observed in magnetars. Therefore, the existence of a common fundamental plane for X-ray emission from pulsars and magnetars in quiescence does not necessarily imply that they possess similar emission mechanisms. It suggests that there is something unknown but common to these two groups of neutron stars.

Further investigations incorporating larger sample sizes, improved error estimations, and new insights into the X-ray emission mechanisms of magnetars and pulsars are required to comprehensively understand the similarities and differences between these two classes of objects. These studies will shed light on the underlying physical processes that shape their X-ray emissions, allowing us to better interpret the observed commonalities and unique features in their emissions. By expanding our knowledge in this area, we can deepen our understanding of the complex nature of neutron stars and advance our understanding of their emission mechanisms.

\subsection{Notes to three special targets}
In our study, there are three sources that warrant further discussion. First, XTE J1810$-$197 stands out as an outlier with its unphysically large power index. The PL-plus-BB model used to describe the X-ray spectrum of XTE J1810$-$197 appears inadequate, as it has been reported that the non-thermal component cannot be constrained in its X-ray spectrum \citep{2004ApJ...605..368G}. Moreover, this source behaves differently compared to other magnetars \citep{2010ApJ...710L.115K,2013MNRAS.434..123V,2023JKAS...56...41S}, indicating the need for alternative spectral models to better describe its emission. A possible alternative model that may better suit the spectrum of XTE J1810$-$197 is two black body components \citep{2018ApJ...852...86M,2021MNRAS.504.5244B}. This approach could potentially offer a more accurate representation of its X-ray emission characteristics.

Regarding Swift J1822.3$-$1606, it was discovered by Swift/BAT as an SGR \citep{2011ATel.3488....1C,2011ApJ...743L..38L}. However, in the $L_{\rm{PL}}$-$T$-$R$ plane, it does not align with the magnetar group but instead falls within the pulsar group. This is further supported by similarities in certain timing properties, such as a weak dipole magnetic field and characteristic age \citep{2011ApJ...743L..38L}. On the other hand, its bright luminosity, which cannot be explained by its low spin-down power, and its bursting activities characterize it as a magnetar \citep{2011ApJ...743L..38L}. Its low surface temperature and old age indicate that Swift J1822.3$-$1606 is likely in a late evolutionary stage. At this stage, magnetars may exhibit behavior more similar to pulsars, as suggested by the similarities between Swift J1822.3$-$1606 and pulsar group. It is possible that both XTE J1810$-$197 and Swift J1822.3$-$1606 are in different stages of evolution or undergoing distinct cooling processes, resulting in their spectral properties deviating from those of typical magnetars. XTE J1810$-$197, for instance, cannot be adequately described by the PL-plus-BB model, while Swift J1822.3$-$1606 shares similarities with pulsars in the $L_{\rm{PL}}$-$T$-$R$ plane.

For the magnetar-like pulsar PSR J1119$-$6127, previous observations and discussions in the literature have revealed that its properties during magnetar-like outbursts, including X-ray bursts, radio emission, and glitches, exhibit remarkable similarities to those of magnetars \citep{2016ApJ...829L..21A,2016ApJ...829L..25G,2018MNRAS.480.3584D,2021MNRAS.503.1214C}. While during its quiescent state, PSR J1119$-$6127 behaves as a normal pulsar with a relatively strong magnetic field \citep{2001ApJ...554..161P,2018MNRAS.480.3584D,2021ApJ...917...56B}. PSR J1119$-$6127 and PSR J1846-0258 are believed to represent a transition between magnetars and pulsars because the observed magnetar-like bursts \citep{2016ApJ...829L..21A,2008Sci...319.1802G}. In our study, the positioning of PSR J1119$-$6127 between magnetars and pulsars in the $L_{\rm{PL}}$-$T$-$R$ plane further suggests its connection or transitional nature between magnetars and pulsars.

\section{Summary}
In this paper, we have presented our finding of the common fundamental plane spanned by power-law luminosity $L_{\rm{PL}}$, black body temperature $T$, black body emitting radius $R$ of magnetars and pulsars. This fundamental plane highlights the significant influence of thermal black body radiation on the determination of their non-thermal power-law luminosity. Despite the distinct non-thermal X-ray emission mechanisms and spectral differences observed between magnetars and pulsars, the existence of this shared fundamental plane suggests a deeper underlying connection between these classes of neutron stars. It underscores the interplay between their thermal emission, non-thermal emission, and timing properties. Further investigations are warranted to enhance our understanding of the magnetospheric emissions from magnetars and pulsars.

Additionally, we discuss three noteworthy sources which may provide valuable insights into the complex relationship between magnetars and pulsars, emphasizing the importance of exploring their unique spectral properties in future research.

\section*{Acknowledgements}
\addcontentsline{toc}{section}{Acknowledgements}
This research has made use of data obtained from the Chandra Data Archive and the Chandra Source Catalog, and software provided by the Chandra X-ray Center (CXC) in the application packages CIAO. We acknowledge the use of public data from the Swift data archive and XMM-Newton Science Archive (XSA). This research has made use of data and software provided by the High Energy Astrophysics Science Archive Research Center (HEASARC), which is a service of the Astrophysics Science Division at NASA/GSFC. This work is supported by the National Science and Technology Council (NSTC) of Taiwan under the grant MOST 111-2112-M-007-019.

\section*{Data availability}
The data underlying this article are available in the article. Chandra data are available through Chandra Data Archive\footnote{\url{https://cxc.harvard.edu/cda/}}. XMM data are available through XMM-Newton Science Archive\footnote{\url{http://nxsa.esac.esa.int/nxsa-web/}}. Swift/XRT data are available through Swift Archive\footnote{\url{https://swift.gsfc.nasa.gov/archive/}}.
\addcontentsline{toc}{section}{Data availability}

\bibliographystyle{mnras}
\bibliography{magnetar.bib}


\bsp	
\label{lastpage}
\end{document}